\documentstyle[12pt]{article}

 \def\be{\begin{equation}}
 \def\ee{\end{equation}}
 \def\bea{\begin{eqnarray}}
 \def\eea{\end{eqnarray}}
 
\begin{document}
\begin{titlepage}
\hfill   OHSTPY-HEP-T-95-007    \\
  \vskip .1in

\center{\Large\bf  Finite Supersymmetric Threshold}
\center{\Large\bf  Corrections to CKM Matrix Elements }
\center{\Large\bf  in the Large $tan\beta$ Regime }

 \vskip .5in

\center{{Tom\'{a}\v{s} Bla\v{z}ek\footnote{On leave of absence from the
Dept. of Theoretical Physics, Comenius Univ., Bratislava, Slovakia} and
Stuart Raby\footnote{Supported in part by DOE
grant -  DOE/ER/01545-646.} }\\{\it Department of Physics, The Ohio State
University}  \\{\it 174 W. 18th Ave., Columbus, OH 43210}\\ {\it
blazek@mps.ohio-state.edu, raby@mps.ohio-state.edu} \\ \vskip .2in  {and} \\
\vskip.2in {Stefan Pokorski\footnote{Supported in part by the Polish Committee
for Scientific Research and by the EU grant ``Flavourdynamnics."}} \\ {\it
Department of
Physics, University of Warsaw }\\
     {\it Warsaw, Poland }\\
{\it stp@dmumpiwh.bitnet}}

\vskip .5in

\flushleft{{\bf Abstract:}}
 We evaluate the finite 1-loop threshold corrections, proportional to
$tan\beta$, to
the down quark mass matrix.  These result in corrections to down quark masses
and to Cabibbo-Kobayashi-Maskawa [CKM] matrix elements. The corrections to
CKM matrix
elements are the novel feature of this paper. For grand unified theories with
large $\tan\beta$ these corrections may significantly alter the low energy
predictions of four of the CKM matrix elements and the Jarlskog parameter J,
a measure of CP violation.
The angles $\alpha,\: \beta$ and $\gamma$ of the unitarity
triangle and the ratio $|{V_{ub} \over V_{cb}}|$, however, are not corrected to
this
order.  We also discuss these corrections in the light of recent models for
fermion
masses. Here the corrections may be useful in selecting among the various
models.
Moreover, if one model fits the data, it will only do so for a particular range
of
SUSY parameters.

\end{titlepage}

\section{Introduction}

\pagestyle{plain}
\pagenumbering{arabic}

Minimal supersymmetric[SUSY] grand unified theories[GUTs] based on the gauge
group SO(10)
require  $tan\beta$ (the ratio of the vacuum expactation values of the two
Higgs scalar
doublets present in the low energy theory) to be of order
$ M_{top}/M_{bottom}\approx 50 $. This follows from the unification of the
top, bottom and tau Yukawa couplings at the GUT scale, $M_{GUT}$, and the
necessity to fit the large top to bottom mass ratio at the weak
scale\cite{als}.
Recent results using a general SO(10) operator analysis for fermion
masses and mixing angles seem to be in significant agreement
with experiment \cite{ADHRS}. It was shown in \cite{HRS,COPW,RH}, however, that
there
are potentially large finite 1-loop corrections (proportional to
tan$\beta$) to the masses of  the down-type quarks at the supersymmetric
threshold.  Note, these corrections were not included in the analysis of
ref.~\cite{ADHRS}. They may be as large as several tens of per cent dependent
on the
sparticle spectrum\cite{COPW}.  Thus they must be included when analyzing any
SUSY theory
with large $\tan\beta$. In this paper we emphasize that the non-diagonal
elements of the
down quark mass matrix also get potentially large corrections; thus leading to
significant corrections to some CKM matrix elements and  the Jarlskog parameter
J. Our
main results are given in equations (\ref{deltami}), (\ref{dvcbub})-
(\ref{dvtstd}) and
(\ref{dJ}) (or in approximate form in eqns. (\ref{dmiapprox}) and
(\ref{approxckm})).
Note, the Cabibbo angle and the CP violating angles $\alpha, \beta$ and
$\gamma$ are not
significantly corrected to this order.

 \section{1-Loop Corrections to the Down Quark Mass Matrix}

When one integrates superpartners out of the minimal supersymmetric standard
model (MSSM) , there are significant
${\cal O}(tan \beta)$ 1-loop corrections to the mass matrix of the
\mbox{down-type} quarks originating in the diagrams with
gluino -- d-type squark and chargino -- u-type squark loops
yielding (see Figures 1a,b,c , for the notation and conventions used
and a short derivation see Appendix)\footnote{We would like to thank Uri Sarid
for bringing our attention to the second chargino diagram , Fig.1c.}
 \be
 {\bf m}_d=(V_d^{\!L\,0})^{\dagger}\:(1\,+\,\epsilon\,{\bf \Gamma}_{\!d}
\,+\,\epsilon\,V_{\!\!{\scriptscriptstyle C\!K\!M}}^{\!0\,\dagger}
%\lambda_u^{\!0Diag}
\,{\bf \Gamma}_{\!u}V_{\!\!{\scriptscriptstyle C\!K\!M}}^{\!0}\,)
\:{\bf m}_d^{\mbox{}_{\!{\scriptstyle 0}}Diag}
\,V_d^{\!R\,0} \; ,
  \label{md}
 \ee
with $\;\;\epsilon=\frac{1}{16\pi^2}\tan\!\beta\;\;\;$   and
 \bea
{\bf \Gamma}_{\!d_{\:{\scriptstyle ij}}}\!&=&\!\!\frac{8}{3}g_3^2\:
(\Gamma_{\!dL}^{\dagger})_{\mbox{}_
{{\scriptstyle i\alpha}}} \:
\int\!\! dk\,\frac{M_{\tilde{g}}} {(k^2+M_{\tilde{g}}^2)
(k^2+m_{\tilde{d}_\alpha}^2)}\:(\Gamma_{\!dR})_{\mbox{}_{{\scriptstyle \alpha
l}}}\:
\frac{({\bf m}_d^{\mbox{}_{\!{\scriptstyle 0}}Diag})^{-1}_
{\mbox{}_{\!{\scriptstyle lj}}}}{\tan\!\beta} \; ,
\label{gammad} \\
{\bf \Gamma}_{\!u_{\:{\scriptstyle ij}}}\!&=&\!\!-\,\lambda_{u_i}^{\!0Diag}\,
(\Gamma_{\!uR}^{\dagger})_{\mbox{}_{{\scriptstyle i\alpha}}}
\:\int\!\! dk\,
\frac{U_{2A}^{*\,\dagger}\:m_{\chi_A}\:V_{\!A2}}{(k^2+m_{\chi_A}^2)
(k^2+m_{\tilde{u}_{\alpha}}^2)}\:(\Gamma_{\!uL})_{\mbox{}_{{\scriptstyle
\alpha j}}}\,(v_u)^{-1} \nonumber \\
&\mbox{}&\!\!+\,g_2\,(\Gamma_{\!uL}^{\dagger})_{\mbox{}_{{\scriptstyle
i\alpha}}}
\:\int\!\! dk\,
\frac{U_{2A}^{*\,\dagger}\:m_{\chi_A}\:V_{\!A1}}{(k^2+m_{\chi_A}^2)
(k^2+m_{\tilde{u}_{\alpha}}^2)}\:(\Gamma_{\!uL})_{\mbox{}_{{\scriptstyle
\alpha j}}}\,(v_u)^{-1}
\;\: .
                                \label{gammau}
 \eea
Uncorrected mass and mixing matrices are labeled by a ``0'' superscript.  The
$6 \times
3$ dimensional matrices $\Gamma_{\!qL}$ and $\Gamma_{\!qR}  \; (q= u, d)$
correspond to the additional transformations necessary to diagonalize squark
mass matrices in a SUSY basis where quark mass matrices are diagonalized.
Expressions for
the ${\bf \Gamma}'s$ are rather complex since they involve  the summation over
the
six-dimensional squark space. It has to be stressed  though, that despite the
explicit
$tan\beta$ term in the denominator  of (\ref{gammad}) and a similar
$(v_u)^{-1}\!=\!(v_d\tan\!\beta)^{-1}$ term  in (\ref{gammau}), there will not
be any
actual  $tan\beta$  suppression in the elements of the ${\bf \Gamma}$'s.  In
the
interaction basis (see the Feynman diagrams Fig.1a,b,c) one can easily
recognize the $R\!-\!L$ mixings among squarks in the loop and the mass
insertions
or mixings on the fermionic line in the diagram. Since each of these
mixings and mass insertions introduces a $\tan\!\beta$ unsuppressed quantity,
the
result represents a $\tan\!\beta$ unsuppressed correction. This correction is
significant since it corrects a $\tan\!\beta$ suppressed mass matrix. To
emphasize
this fact the large Higgs vev ratio was pulled out into the $\epsilon$'s in
(\ref{md}).
As a net effect, one can expect (at least some) terms in the ${\bf \Gamma}'s$
to be of
order $(0.1 - 1)$.   These terms are then enhanced by a factor of $\tan\!\beta$
(multiplying a standard small loop
factor $(16\pi^2)^{-1}$ in our definition of $\epsilon$)  and thus lead to
significant mass matrix corrections.
\par
Concentrating on the above mentionned diagrams one has to mention that
there are also neutralino diagrams which contribute by finite
${\cal O}(tan \beta)$ terms to the $d$ quark mass matrix. However, we have
checked
that (assuming degenerate gauginos at $M_{GUT}$) these contributions are less
than the leading gluino corrections
roughly by a factor 16, as a result of smaller couplings, gaugino masses
and group factors. Therefore these diagrams will not be discussed separately
in this paper although they are included in our numerical analysis in section 4
where we discuss their effects.
\par
In order to gain some intuition for the ${\bf \Gamma}'s$ one can find an
explicit form
for them in the following approximation. First, neglect the second chargino
diagram
(Fig.1c), since it is
suppressed by a smaller coupling constant compared to the diagrams
in Fig.1a and Fig.1b. Then in the evaluation of the remaining
two diagrams assume that squark mass matrices are diagonalized
in generation space by the same rotations as the corresponding quark matrices.
This approximation is valid assuming universal scalar masses and trilinear
scalar interactions proportional to Yukawa interactions at the low energy SUSY
scale.

That means that in this approximation the
matrices $\Gamma_{\!\!qL}$ and $\Gamma_{\!\!qR}$, ($q=d,u$) are diagonal in
generation
space and are not completely trivial only because of the mixing between squarks
of the same generation.
The integrals in (\ref{gammad}) and (\ref{gammau}) are then easy to do along
with the summation over $\alpha = 1, \cdots, 6$ , i.e. over the squark mass
eigenstates.
The  ${\bf \Gamma}'s$ are then proportional to the off-diagonal term
of the down (up) squark mass matrix for each individual generation separately.
We find
\bea
{\bf \Gamma}_{\!d_{\:{\scriptstyle
ij}}}&=&\frac{8}{3}g_3^2\,M_{\tilde{g}}\,\mu\,
I_3(M_{\tilde{g}}^2,m_{\tilde{d}_{i_1}}^2,m_{\tilde{d}_{i_2}}^2)\,\delta_{ij}
\;\, , \label{gammadapprox} \\
{\bf \Gamma}_{\!u_{\:{\scriptstyle ij}}}&=&U_{2A}^{*\,\dagger}\,m_{\chi_A}\,
V_{\!A2}\,A_0\,I_3(m_{\chi_A}^2,m_{\tilde{u}_{i_1}}^2,m_{\tilde{u}_{i_2}}^2)
\,(\lambda_u^{\!0Diag})_{ij}^2  \;     ,\label{gammauapprox}
\eea
with the function $I_3$ given by $$ I_3(a,b,c) = {ab \ln({a \over b}) + bc
\ln({b \over
c}) + ac \ln({c \over a}) \over (a - b)(b - c)(a - c)} .$$
 Terms suppressed by $\tan\!\beta$ have been neglected
in these expressions.
 In this approximation both ${\bf \Gamma }$ matrices are diagonal which makes
calculations of the corrections to the masses and mixing angles
in terms of mass eigenstates simple. Besides that, note the large hierarchy
in ${\bf \Gamma}_{\!u}$, and a much milder hierarchy in ${\bf \Gamma}_{\!d}$
based just on the non-equality of the  squark masses.
 Note, if ${\bf \Gamma}_{\!d}$ were completely
proportional to the identity matrix (i.e. the case of complete squark
degeneracy), the
gluino loop would not contribute to quark mixing corrections at all
\footnote{Also note that the same analysis could be done for
the corrections to the up quark mass matrix and the above mentioned
approximation would show that the relevant ${\bf \Gamma }$ matrices
(analogous to (\ref{gammadapprox}) and (\ref{gammauapprox}) )
become suppressed by $\tan\!\beta$ after the up (instead of down)
quark mass matrix is pulled out of the expression analogous to (\ref{md}).
Thus in this case there are no corrections proportional to $\tan\beta$.
There are however corrections to charged lepton masses proportional to
$\tan\beta$.  These
are  smaller than those for down quarks but are still significant and must be
included in
any fermion mass analysis.}. \par
One knows though, that the approximation used to
derive (\ref{gammadapprox}) and (\ref{gammauapprox}) is not correct.
The initial conditions at $M_{GUT}$ need not be universal and, even if they
were, squark masses and trilinear couplings run between the
GUT (or string) and the low energy SUSY scales and violate our assumptions. As
a result the explicit form of the potentially significant
(i.e. $\tan\!\beta $ unsuppressed) elements in the ${\bf \Gamma }$'s is clouded
by the fact that they no longer remain diagonal in generation space.
In order to evaluate these effects we have performed a numerical analysis.
The results are found in Section 4.  We also show that our naive approximation,
equations
(\ref{gammadapprox}, \ref{gammauapprox}), when suitably modified to take into
account non-universal squark masses gives results which agree to within 25\%
with the two-loop numerical analysis.
\par
In order to figure out the explicit form of the 1-loop threshold
corrections  to the CKM matrix elements as well as to quark masses in terms of
the ${\bf \Gamma }$ matrix elements
one can define an unknown hermitian matrix ${\bf B}$ as
 \be
 V_d^L=(1+i\epsilon{\bf B})\,V^{L\,0}_d
  \label{Vd}
 \ee
 where, again, $V^{L\,0}_d$ is the matrix diagonalizing down quarks in the
absence of
the SUSY corrections. Since there are no large
( i.e. ${\cal O}(\tan\!\beta$) ) corrections to the up quark mass matrix
 \be
 V_{CKM}\equiv
V_u^LV_d^{\!L\,\dagger}=V_u^{\!L\,0}(V_d^{\!L\,0})^{\dagger}(1-i\epsilon{\bf
B})=V_{CKM}^0(1-i\epsilon{\bf B})\;.
  \label{CKM}
 \ee
{\bf B} is determined through the diagonalization condition
 \be
 ({\bf m}_d^{\bf Diag})^2\equiv Diag(m_{d_1}^2,m_{d_2}^2,m_{d_3}^2)
=V_d^L\:{\bf m}_d{\bf m}_d^{\dagger}\:V_d^{\!L\,\dagger}\;,
 \label{6}
 \ee
 where both $V_d^L$ and ${\bf m}_d$ on the r.h.s. are to be expanded to first
order in $\epsilon$ according to (\ref{Vd}) and (\ref{md}).

\subsection{Corrections to Down Quark Masses}

Diagonal elements of this matrix equation (\ref{6}) specify the corrections to
the
masses of the $d$, $s$ and $b$  ($d_1$, $d_2$ and $d_3$)  quarks.
Note that the terms containing unknown ${\bf B}$ elements drop out of
these equations :
 \be
\frac{\delta m_{d_i}}{m_{d_i}}=\epsilon\,{\rm Re}({\bf \Gamma}_{\!d})_
{\mbox{}_{{\scriptstyle ii}}}+\epsilon\,
[V_{\!\!{\scriptscriptstyle C\!K\!M}}^{\!0\,\dagger}
\,{\rm Re}({\bf \Gamma}_{\!u}) \, V_{\!\!{\scriptscriptstyle C\!K\!M}}^{\!0}]_
{\mbox{}_{{\scriptstyle ii}}}  \;.
                                  \label{deltami}
 \ee
This is an exact formula where the effects of squark
rotations are fully included in the ${\bf \Gamma }$'s.
Since the ${\bf \Gamma_u }$ matrix has some generation hierarchy
(for more discussion on this see Section 4)
due to the Yukawa couplings in the chargino loop
the dominant correction from the chargino diagram
goes to the $b$ quark mass correction:
\be
\left(\frac{\delta m_b}{m_b}\right)_{\!\chi^+}=\epsilon\,{\rm Re}
({\bf \Gamma}_{\!u})_{\mbox{}_{{\scriptstyle 33}}}+\epsilon{\cal O}(10^{-3})
\;.\label{charg}
\ee
The suppression in the second term above is caused by the
hierarchies present in (\ref{deltami}).  The largest next-to-leading correction
 indicated above results, for example, from the term
$(V_{\!\!{\scriptscriptstyle C\!K\!M}}^{\!0\,\dagger})_{32}
Re({\bf \Gamma}_{\!u})_{23}\,
(V_{\!\!{\scriptscriptstyle C\!K\!M}}^{\!0})_{33}$ where
two orders come from $V^*_{cb}$ and at least one order from ${\bf \Gamma }_
{\!u_{{\scriptstyle 32}}}$.

\par
Note that the corrections to the masses of the $s$ and $d$ quarks
can easily be as significant, or even larger than the correction to the $b$
quark
mass. While the gluino correction (which is the largest correction
to each quark mass) to the $b$ quark mass is larger
due to the smaller $b$ squark masses (in a universal-like scenario where
one starts with all soft squark masses equal at the GUT scale),
the chargino correction may invert the net effect since it is always
of opposite sign to the gluino correction and its contribution to the two
lighter quarks is small.
\subsection{Corrections to CKM Matrix Elements}

The non-diagonal equations, i.e. those with zeros on the l.h.s. of the matrix
equation (\ref{6}), lead to
\be
-i\epsilon{\bf B}_{ij}=\epsilon\,{\bf \Gamma}_{\!d_{\:{\scriptstyle ij}}}+
\epsilon\,(\,V_{\!\!{\scriptscriptstyle C\!K\!M}}^{\!0\,\dagger}
{\bf \Gamma }_{\!u}\,V_{\!\!{\scriptscriptstyle C\!K\!M}}^{\!0}\,)_
{\mbox{}_{\!{\scriptstyle ij}}}\:(1+{\cal O}(\frac{m_{d_i}^2}{m_{d_j}^2}))\;,
\ee
where $ij$ indices correspond to the $12$, $13$ or $23$ combinations and the
transposed elements ( for $i>j$ ) are obtained by the hermiticity of {\bf B}.
 The diagonal elements of {\bf B} remain undetermined by this procedure but
to the first
order in the $\epsilon $-expansion they can be removed by phase redefinitions
of the
$b$, $s$ and $d$ fields. We thus set the diagonal elements of {\bf B} to zero.
 \par
 Then from (\ref{CKM}) we can easily derive \footnote{Zero superscripts are
dropped from now on since they make no difference in the following
expressions.}
 \bea
\delta V_{cb}=\!\!\!\!\!\!&\mbox{}&\!\!\!\!\epsilon\,[\,V_{cd}\,
{\bf \Gamma }_{\!d_{\:{\scriptstyle 13}}}+
V_{cs}\,{\bf \Gamma }_{\!d_{\:{\scriptstyle 23}}}\,] \hspace{6 cm} \nonumber \\
\mbox{}&+&\!\!\!\!\epsilon\,[\: \{ \delta_{2j}-
(V_{\!{\scriptscriptstyle C\!K\!M}})_{\mbox{}_{\!{\scriptstyle 23}}}
(V_{\!{\scriptscriptstyle C\!K\!M}}^{\dagger})_{\mbox{}_{\!{\scriptstyle
3j}}}\}
\,{\bf \Gamma }_{\!u_{\:{\scriptstyle jk}}}\,
(V_{\!{\scriptscriptstyle C\!K\!M}})_{\mbox{}_{\!{\scriptstyle k3}}}\,]
\hspace{2 cm}
\label{dvcb}
\\ \vspace{1 mm}
\delta V_{ub}=\!\!\!\!\!\!&\mbox{}&\!\!\!\!\epsilon\,[\,V_{ud}\,{\bf \Gamma
}_{\!d_{\:{\scriptstyle 13}}}+
V_{us}\,{\bf \Gamma }_{\!d_{\:{\scriptstyle 23}}}\,] \hspace{6 cm} \nonumber \\
\mbox{}&+&\!\!\!\!\epsilon\,[\: \{ \delta_{1j}-
(V_{\!{\scriptscriptstyle C\!K\!M}})_{\mbox{}_{\!{\scriptstyle 13}}}(V_{\!
{\scriptscriptstyle C\!K\!M}}^{\dagger})_{\mbox{}_{\!{\scriptstyle 3j}}} \}
\,{\bf \Gamma }_{\!u_{\:{\scriptstyle jk}}}\,
(V_{\!{\scriptscriptstyle C\!K\!M}})_{\mbox{}_{\!{\scriptstyle k3}}}\,]
\hspace{2 cm}
\label{dvub}
\\ \vspace{1 mm}
\delta V_{td}=\!\!\!\!&-&\!\!\!\!\epsilon\,[\,V_{ts}\,{\bf \Gamma }_{\!d_{\:
{\scriptstyle 21}}}^{\dagger}+V_{tb}\,{\bf \Gamma }_
{\!d_{\:{\scriptstyle 31}}}^{\dagger}\,] \hspace{6 cm}\nonumber \\
\mbox{}&-&\!\!\!\!\epsilon\,[\: \{ \delta_{3j}-
(V_{\!{\scriptscriptstyle C\!K\!M}})_{\mbox{}_{\!{\scriptstyle 31}}}
(V_{\!{\scriptscriptstyle C\!K\!M}}^{\dagger})_{\mbox{}_{\!{\scriptstyle 1j}}}
\}
\,{\bf \Gamma }_{\!u_{\:{\scriptstyle jk}}}^{\dagger}\,
(V_{\!{\scriptscriptstyle C\!K\!M}})_{\mbox{}_{\!{\scriptstyle k1}}}\,]
\hspace{2 cm}
\label{dvtd}
\\ \vspace{1 mm}
\delta V_{ts}=\!\!\!\!\!\!&\mbox{}&\!\!\!\!\epsilon\,[\,V_{td}\,{\bf \Gamma
}_{\!d_{\:{\scriptstyle 12}}}-
V_{tb}\,{\bf \Gamma }_{\!d_{\:{\scriptstyle 32}}}^{\dagger}\,]\,+\,\epsilon\,
[\,(V_{\!{\scriptscriptstyle C\!K\!M}})_{\mbox{}_{\!{\scriptstyle 31}}}
(V_{\!{\scriptscriptstyle C\!K\!M}}^{\dagger})_{\mbox{}_{\!{\scriptstyle 1j}}}
\,{\bf \Gamma }_{\!u_{\:{\scriptstyle jk}}}\nonumber \\
&\mbox{}&\;\;\;\;-(V_{\!{\scriptscriptstyle C\!K\!M}})_
{\mbox{}_{\!{\scriptstyle 33}}}(V_{\!{\scriptscriptstyle C\!K\!M}}^{\dagger})_
{\mbox{}_{\!{\scriptstyle 3j}}}\,
{\bf \Gamma }_{\!u_{\:{\scriptstyle jk}}}^{\dagger}]\,
(V_{\!{\scriptscriptstyle C\!K\!M}})_{\mbox{}_{\!{\scriptstyle k2}}}\;\:.
\label{dvts}
 \eea
As already mentioned, the numerical analysis shows that both
${\bf \Gamma }_{\!u}$ and ${\bf \Gamma }_{\!d}$ matrices keep track
of the generation hierarchy from the Yukawa sector with the
$33$ element of the order of 0.1-1 and the relevant $12$, $13$ and $23$
elements of small magnitude.
Together with the hierarchy in the CKM matrix this implies that in each of
the previous equations the dominant correction is the one containing
the ${\bf \Gamma}_{\!u_{\,{\scriptstyle 33}}}$ term.  However the corrections
due to the other terms are non-neglible resulting in a 25\% effect.  Thus a
good approximation to the exact results of equations (\ref{dvcb}) -
(\ref{dvts}) is given by (for more detail, see Section 4)
\be
 \frac{\delta V_{cb}}{V_{cb}}\approx -
\epsilon\,[{\bf \Gamma }_{\!u_{\,{\scriptstyle 33}}} - {\bf \Gamma
}_{\!u_{\,{\scriptstyle 22}}} - \frac{V_{cs}}{V_{cb}}\; {\bf \Gamma
}_{\!d_{\,{\scriptstyle 23}}}] \equiv - \epsilon \;\Delta \label{dvcbub}
\ee
\be
 \frac{\delta V_{cb}}{V_{cb}} \approx \frac{\delta V_{ub}}{V_{ub}} \approx -
\epsilon \; \Delta \label{dvcbub2}
 \ee
 \be
 \frac{\delta V_{ts}}{V_{ts}}\approx\frac{\delta V_{td}}{V_{td}}\approx -
\epsilon \; \Delta^*.
\label{dvtstd}
\ee

The results of equations (\ref{dvcbub2}) and (\ref{dvtstd}) follow directly
from the unitarity of the CKM matrix and the fact that these are the only terms
which receive significant corrections.

{\em Note that as a consequence the ratio
$V_{ub}/V_{cb}$ remains unchanged.} In addition the numerical analysis shows
that ${\rm
Re}{\bf \Gamma}_{\!u_{\,{\scriptstyle 33}}} \gg {\rm Im}{\bf
\Gamma}_{\!u_{\,{\scriptstyle 33}}}$ thus these  dominant corrections to the
CKM
elements are equal in magnitude, but opposite  in sign, to the chargino
corrections to
the $b$ quark mass, eq.(\ref{charg}).

\par
The other five CKM elements get the corrections of the form similar to
(\ref{dvcb}) - (\ref{dvts}). However, large ${\bf \Gamma }$ elements are
always in the product with a small CKM matrix
element, and the terms containing large diagonal CKM matrix
elements are in the same way pushed down by small ${\bf \Gamma }$ elements
in these corrections. Hence
the actual numerical values of the corrections to $V_{ud}$, $V_{us}$,
$V_{cd}$, $V_{cs}$ and $V_{tb}$ are not significant, at least not at the
present level of
experimental accuracy. As an example, the dominant correction to, let's say
$V_{us}$ goes
like  \be
 \delta V_{us}\sim \epsilon V_{ub}V_{ts}V_{tb}^*
{\bf \Gamma }_{\!u_{\:{\scriptstyle 33}}}^* < 0.001 \;. \label{dvcs}
\ee

 \section{CP Violating Parameters}
The Jarskog parameter which measures CP violation can be obtained from the four
CKM-matrix elements left after crossing out any row and any column of this
matrix\cite{CJ}:
\be
 J \sum_{\gamma,\l}\epsilon_{\alpha\beta\gamma}\epsilon_{jkl}=\,Im\,[\,
 V_{\alpha j}V_{\beta k}V_{\alpha k}^*V_{\beta j}^*]\:.
 \label{J}
\ee
Consider the product
\be
J \,=\,Im\,[\,V_{cs}V_{tb}V_{cb}^*V_{ts}^*]
\ee
Using the formula (\ref{dvcbub}) and (\ref{dvtstd}) from the previous section
it is easy to obtain the leading correction
\be
\delta J \approx - 2 \epsilon\,{\rm Re}[\Delta]\, J
\label{dJ}
\ee
{\em This threshold correction to J may significantly alter the prediction for
$\epsilon_K$ in SUSY GUT models with large $tan\beta$.}
\par
Note, it is not obvious how this result is obtained for other equivalent
definitions
of J. For example, at first glance one might guess that $\delta J\approx\,0$
for J defined by  \mbox{$\:J =\,{\rm Im}\,[\,V_{ud}V_{cs}V_{us}^*V_{cd}^*]\:$}
. However, such a guess does not take into account that we have the imaginary
part of the product in (\ref{J}) and imaginary parts are small for every CKM
matrix element, even if its absolute value is close to one.
In this case the small corrections to the large CKM matrix elements become
important and, in fact,
it is corrections to $V_{cs}$ and $V_{us}$ that lead to the result
(\ref{dJ}).

 \par
 Finally, we note that although J changes, {\em the angles of the unitarity
triangle
 remain uncorrected to this order.}  This is easily understood from a
geometrical point of view. For the
``standard''  choice of its sides -- $\mid V_{ud}V_{ub}^*\mid $,
$\mid V_{cd}V_{cb}^*\mid $ and $\mid V_{td}V_{tb}^*\mid $ -- each side contains
 one element which gets a significant correction and (as a consequence of the
unitarity of the CKM matrix discussed earlier) these corrections are
identical in magnitude (see (\ref{dvcbub}) and (\ref{dvtstd})). Hence, the
sides are contracted (or stretched) by the same multiplicative factor and the
angles stay the same. The area of the triangle gets corrected, of course, twice
 as much as the sides, and that is the reason for the factor of two in
(\ref{dJ}) (recall that J measures the area of the triangle).

 \section{Numerical Analysis and Conclusions}

In our numerical analysis we took the initial conditions (values at the
GUT scale) for the dimensionless couplings from the
SO(10) models \cite{ADHRS} which give predictions for the low
energy data in good agreement with experiment. The initial values
for the dimensionful soft SUSY breaking parameters were taken from
ref.\cite{COPW,OP} in order to guarantee the radiative electroweak symmetry
breaking at the weak scale. We focused mainly on simple non-universal cases.
The numerical results presented below were obtained for
$m^2_{H_1}\!=\!2.0m_0^2$, $m^2_{H_2}\!=\!1.5m_0^2$ and
all other scalar masses equal $m_0^2$.
Next, we used 2-loop renormalization group equations\cite{MV}
 to run all the couplings and mass parameters to the low energy
scale. Leading corrections to the CKM matrix elements have appeared practically
 independent of the exact value of the low energy SUSY scale between
$M_Z$ and 500 GeV (changes were within 1\% of the mass or the CKM element in
question).
In the actual numerical analysis
the ${\bf \Gamma}$-matrices have been evaluated according to the following
formulae (note that there are no divergent pieces from the integrals in
(\ref{gammad}) and (\ref{gammau}) and that the chargino summation is easy to
do) :
 \bea
{\bf \Gamma}_{\!d_{\:{\scriptstyle ij}}}\!\!\!&=&\!\!\frac{8}{3}g_3^2\,
(\Gamma_{\!dL}^{\dagger})_{\mbox{}_{\!{\scriptstyle i\alpha}}}
\:\frac{-m^2_{\tilde{d}_{\alpha}}}{m^2_{\tilde{d}_{\alpha}}-M^2_{\tilde{g}}}
\,\ln\frac{m^2_{\tilde{d}_{\alpha}}}{M^2_{\tilde{g}}}\:(\Gamma_{\!dR})_
{\mbox{}_{\!{\scriptstyle \alpha j}}}
\:\frac{M_{\tilde{g}}}{m_{d_j}\tan\!\beta}\\
{\bf \Gamma}_{\!u_{\:{\scriptstyle ij}}}\!\!\!&=&\!\!\lambda_{u_i}\,
(\Gamma_{\!uR}^{\dagger})_{\mbox{}_{\!{\scriptstyle i\alpha}}}
\:(m^2_{\tilde{u}_{\alpha}}-\mid\! M_2\!\mid^2)\:
I_3(m^2_{\chi_1},m^2_{\chi_2},m^2_{\tilde{u}_{\alpha}})\:(\Gamma_{\!uL})_
{\mbox{}_{\!{\scriptstyle \alpha j}}}\,\frac{\mu}{v_u} \nonumber\\
&\mbox{}&\!\!\!\!\!-g_2^2
(\Gamma_{\!dL}^{\dagger})_{\mbox{}_{\!{\scriptstyle i\alpha}}}
I_3(m^2_{\chi_1},m^2_{\chi_2},m^2_{\tilde{u}_{\alpha}})\:(\Gamma_{\!uL})_
{\mbox{}_{\!{\scriptstyle \alpha j}}}\,M_2\mu\,.
 \eea
$M_2$ is the wino mass parameter and terms suppressed by $\tan\!\beta$ were
dropped. The summation is only over $\alpha=1,...6$
 (there's no summation over $i,j$ on the r.h.s.\  of these equations). This
 summation could be done analytically in terms of the mass eigenvalues,
however the expressions are long and don't provide much insight, so we keep
rather the compact forms above.
\par
Typical values for these matrices at the weak scale follow -
\bea
\epsilon{\bf \Gamma }_{\!d}\!\!\!\!&\approx&\!\!\!\!\!\left(\!\!\!
\begin{array}{ccc}
0.180+i\,3\!\cdot\!10^{-8} & -3\!\cdot\!10^{-6}-i\,9\!\cdot\!10^{-7} &
8\!\cdot\!10^{-5}+i\,2\!\cdot\!10^{-5}  \\
-3\!\cdot\!10^{-6}+i\,7\!\cdot\!10^{-7}  & 0.180-i\,4\!\cdot\!10^{-8} &
-4\!\cdot\!10^{-4}+i\,6\!\cdot\!10^{-6}  \\
-9\!\cdot\!10^{-5}+i\,8\!\cdot\!10^{-5} &
3\!\cdot\!10^{-5}+i\,1\!\cdot\!10^{-5}  &
0.218-i\,3\!\cdot\!10^{-9} \end{array}
\!\!\!\right) \nonumber\\
\epsilon{\bf \Gamma }_{\!u}\!\!\!\!&\approx&\!\!\!\!\!\left(\!\!\!
\begin{array}{ccc}
-0.022+i\,5\!\cdot\!10^{-16} &-9\!\cdot\!10^{-7}+i\,3\!\cdot\!10^{-7}&
-3\!\cdot\!10^{-6}+i\,6\!\cdot\!10^{-6}  \\
-9\!\cdot\!10^{-7}-i\,3\!\cdot\!10^{-7}  & -0.022-i\,3\!\cdot\!10^{-11}  &
-1\!\cdot\!10^{-4}+i\,5\!\cdot\!10^{-9}  \\
1\!\cdot\!10^{-5}+i\,3\!\cdot\!10^{-5}  &
5\!\cdot\!10^{-4}-i\,1\!\cdot\!10^{-8}  &
-0.116+i\,7\!\cdot\!10^{-10}
\end{array} \!\!\! \right)\nonumber
\eea
where in this case we used the GUT scale values, $M_{1/2}\!=\!400GeV$,
$m_0\!=\!250GeV$,
and  $A_0\!=\!-1100GeV$, the weak scale value, $\mu\!=\!270GeV$,  and Model 4
of
\cite{ADHRS} for the Yukawa matrices (with the weak scale values $\lambda_t
=1.01, \tan\beta =
53$ and $V_{cb}^0 = 0.038$ as output).  With these inputs we find
$M_{\tilde{g}} = 1029 GeV, A_t = (A_u)_{33} = - (736 +
i 6\!\cdot\!10^{-7}) GeV$, up-squark mass eigenvalues (in GeV) $(976, \; 976,
\; 951, \;
951, \; 869,$ $695)$ and down-squark mass eigenvalues (in GeV) $(980, \; 979,
\; 949,
\; 948, \; 820,$ $ 757)$.
To gain some intuition for the size of the
corrections, these particular values  lead to $\delta
m_b/m_b\!=\!10.2\%\,,\;\delta
m_s/m_s\!=\!15.7\%\,,\; \delta m_d/m_d\!=\!15.7\%\,,\;$ $\delta
V_{cb}/V_{cb}\!=\!\delta
V_{ub}/V_{ub}\!=\!\delta V_{ts}/V_{ts} =\delta V_{td}/V_{td}=8.0\%$ and $\delta
J/J\!=\!16.6\%$. As we discussed earlier the approximation of retaining only
the $ {\bf \Gamma }_{\!u_{\,{\scriptstyle 33}}}$ term in equation(\ref{dvcbub})
 does not work extremely well since it predicts an 11.6\%
correction. However, this leading correction is then lowered by about 1.5\%
coming from the ${\bf \Gamma_d}$ term in (\ref{dvcb})-(\ref{dvts}) and by
additional 2\% from the subleading ${\bf \Gamma_u}$ terms.  $\;V_{ud},\:
V_{us},\:V_{cd},\:V_{cs}$ and $V_{tb}$ get a relative
correction less than 1\%, e.g. $\delta V_{us}/V_{us}\!=\!0.01\%$. Similarly,
the corrections to the angles $\alpha,\:\beta$ and
$\gamma$ of the  unitarity triangle are much below 1\%. Neutralino corrections
have been included in the above numerical analysis.  Their effects are as
follows: the $b$ mass is reduced
by 1.6\% and the masses of $s$ and $d$ are reduced by 1.3\%. Integrating out
neutralinos has less than a 1\% impact on the CKM elements and CP violating
parameter, J.
\par
We would like to emphasize that such
corrections  are generic for a large subspace of the  allowed parameter space.

\subsection{Approximate Formulae for Mass and Mixing Angle Corrections} In
eqns.
(\ref{gammadapprox}) and (\ref{gammauapprox}), we presented the results of a
naive
approximation which assumes that squark mass matrices are diagonalized
in generation space by the same rotations as the corresponding quark matrices.
This approximation is valid in the case of universal scalar masses and
trilinear scalar interactions proportional to Yukawa interactions when, in
addition, one also neglects the renomalization group running from $M_G$ to the
low energy SUSY scale.  If one now includes the effect of RG
running, quark and squark mass matrices can no longer be diagonalized in
generation space by the same unitary transformations and the A parameters are
no longer
universal. We have checked that a simple approximation for the corrections to
down quark
masses and the CKM matrix elements (valid to 25\%) can be obtained by using the
results of eqns.
(\ref{gammadapprox}) and (\ref{gammauapprox}) with the values of squark and
gluino masses
obtained by RG running as input and by replacing $A_0$ with $A_t$ (for the
third
generation) and the chargino mass with the low energy value of $\mu$. This
approximation
has been widely used in the  previous papers\cite{HRS,RH,COPW,OP,BOP} where
large bottom
mass corrections have been  recognized. In particular, in this improved
approximation
 \bea
 \delta m_d&\approx&(\,\tilde{\epsilon_d}_1+\tilde{\epsilon_d}_2\,
{\cal O}(10^{-6})\,)\:m_d \nonumber\\
 \delta m_s&\approx&(\,\tilde{\epsilon_s}_1+\tilde{\epsilon_s}_2\,
{\cal O}(10^{-4})\,)\:m_s \label{dmiapprox}\\
 \delta m_b&\approx&(\,\tilde{\epsilon_b}_1+\tilde{\epsilon_b}_2\,(\,|V_{tb}|^2
+{\cal O}(10^{-6})))\:m_b \;,\nonumber
 \eea
where
\bea
\tilde{\epsilon_{d_i}}_1&=&\frac{2\alpha_s}{3\pi}\,\mu M_{\tilde{g}}
I_3(M_{\tilde{g}}^2,m_{\tilde{d}_{i_1}}^2,m_{\tilde{d}_{i_2}}^2)\,\tan\!\beta
\\
\tilde{\epsilon_{d_i}}_2&=&\frac{1}{16\pi^2}\,\mu A_{u_i} \lambda_{u_i}^2
I_3(\mu^2,m_{\tilde{u}_{i_1}}^2,m_{\tilde{u}_{i_2}}^2) \,\tan\!\beta .
\label{epstilde}
\eea

The analogous corrections for CKM matrix elements, also valid to about 25\%,
are given by
\be
 \frac{\delta V_{cb}}{V_{cb}}\approx\frac{\delta V_{ub}}{V_{ub}}\approx
 \frac{\delta V_{ts}}{V_{ts}}\approx\frac{\delta V_{td}}{V_{td}}\approx
-\tilde{\epsilon_b}_2, \label{approxckm}
\ee
where $\tilde{\epsilon_b}_2$ is defined in the eq.(\ref{epstilde}).
\par
An important feature of the $b$ quark mass correction is that
the gluino and chargino contributions are of the opposite signs and thus there
is
a partial cancellation between them.  This effect with
its consequences has been carefully studied in \cite{COPW,OP,BOP}.
In these papers it was shown that the magnitude of the gluino contribution
is always two to three times larger than the chargino contribution and
can be as large as 50\% for universal scalar masses at $M_G$.  For
non-universal scalar
masses the corrections can be smaller.

\subsection{Consequences for Models of Fermion Masses}

It is interesting to see what effect these corrections have for recent
models of fermion masses and mixing angles.  In the model of ref.\cite{dhr} the
 value of
$|\,V_{cb}|$ is of order .054.  This is large compared to the latest
experimental
values. In this model, $\tan\!\beta$ can be either small or large.   We would
have to be in the large $\tan\!\beta$ regime for these corrections to be
significant.
In addition consider models 4,6 and 9 of ref.\cite{ADHRS}. In these models
$\tan\!\beta$ is expected to be large.  Recall that the model independent
experimental value of $|\,V_{cb}|$ is  $0.040\pm 0.003$ according to \cite{N}
or $0.040\pm 0.005$ based on \cite{PP}. For models 6 and 9, the predicted value
of $V_{cb} \sim .048 - .052$ is at the upper end of the experimentally allowed
range.  For all these models we would choose $\mu M_{\tilde{g}}\!<\!0$ so that
the chargino correction to the $b$ quark mass is positive and hence
$\delta V_{cb}/V_{cb}\!<\!0$. As a consequence the gluino correction is
negative which gives $\delta m_b\!<\!0$. This has the effect of decreasing the
prediction for $m_t$, since a smaller top Yukawa coupling is now needed to
fit the experimental ratio  $m_b/m_{\tau}$.  These corrections apparently
improve the predictions of the above models. However, the corrections to the
strange and down quark masses, which are equal and negative, may be a problem
since both ratios $m_u/m_d$ and $m_s/m_d$ were rather large and now
the first one gets even bigger while the second one stays the same.
This problem is exacerbated by the fact that the authors in \cite{BOP} find no
solutions
for  $|{\delta m_b \over m_b}| < 10\%$ with  $\mu M_{\tilde{g}}<0$   consistent
with both
the experimental rate for $b\rightarrow s\gamma$  and the cosmological
constraint on the energy density of the Universe.  There are solutions for
larger
values of $|{\delta m_b \over m_b}|$ but this range of parameters may seriously
be
constrained by the ratios $m_u/m_d$ and $m_s/m_d$.

For model 4 of ref.\cite{ADHRS} however the situation may be better. In this
model
$|\,V_{cb}|$ is acceptably small ($V_{cb} \sim .038 - .044$).  However $J$ is
too  small
and thus the bag constant, $B_K$, needed to fit $\epsilon_K$ is too  large,
i.e. greater
than 1.  In this case we need  $\delta J/J\!>\!0$. This would also increase
$|\,V_{cb}|$ by half as much, which may be acceptable.  In this case the
chargino
correction to the $b$ quark mass is negative.  Thus the gluino correction to
$m_b$ is
positive and $\delta m_b\!>\! 0$.   As a result the top quark mass prediction
increases.
This restricts  the magnitude of the effect to values of  $|{\delta m_b \over
m_b}| <
10\%$. In this case both $m_s$ and $m_d$ increase, which improves the agreement
with
experiment in  the $m_s/m_d$ -- $m_u/m_d$ plane. Finally the $b\rightarrow
s\gamma$ decay rate and the cosmological constraint can be satisfied\cite{BOP}.

 Note that in either scenario the angles $\alpha, \; \beta$ and $\gamma$ of the
unitarity triangle and the ratio  $V_{ub}/V_{cb}$ remain unchanged.  These
correlations
of quark mass and mixing angle predictions with the sign of $\mu M_{\tilde{g}}$
are very
intriguing,  especially since this sign may be determined independently once
SUSY
particles  are observed. In a particular model the allowed maximal corrections
to masses
and mixing angles may represent new constraints on the magnitude and sign of
the SUSY parameters.
\par
In summary, finite SUSY corrections to the masses of the down-type quarks may
be significant in the limit of large $\tan\!\beta$. In this paper we have
shown that the CKM matrix elements $V_{cb}, V_{ub}, V_{ts}$ and $V_{td}$
receive similar corrections, while the correction to the Jarlskog parameter is
enhanced
by a factor of two. The other elements of the CKM matrix and the angles of the
unitarity triangle receive only small corrections, down by a factor
$\tan\!\beta$
or suppressed by the generation hierarchy present in Yukawa, CKM or
${\bf \Gamma}$ matrices.

 \section{Appendix}
 \vspace{4 mm}
 \par
Conventions of
the Standard Model are fixed by ${\cal L}_{Yukawa}$=\mbox{$H_q\,\bar{Q}_L\,
{\bf \lambda_q}\,q_R$ ,} quark mass matrix rotations by
$m_q^{Diag}=V_q^L\,m_q\,V_q^{\!R\,\dagger}$ and the CKM
matrix is defined as $V_{CKM}=V_u^LV_d^{\!L\,\dagger}$.
In the MSSM the relevant term in the superpotential
is then $W=\hat{\bar{q}}\,{\bf \lambda_q^{\dagger}}\hat{Q}\hat{H}_q$.
\par
Looking closely at the SUSY threshold corrections to the $d$ quark masses
there are the following 1-loop diagrams contributing significantly in large
$\tan\!\beta$ limit.
\vspace{3 mm}
\par
 $ i) \:\;gluino \;diagram $
 \vspace{3 mm}
 \\
Using Dirac notation the quark-squark-gluino interaction, relevant for this
paper, reads
\be
{\cal
L}_{int}=-\sqrt{2}g_3\,(\frac{\lambda^A}{2})_{ab}\{+(\bar{d}_aP_R\tilde{g}^A)\tilde{d}_{L\,b}-\tilde{d}_{R\,a}^{\dagger}(\bar{\tilde{g}}^AP_Rd_b)\}+h.c.\,.
\ee
The $squark$ interaction eigenstates are turned into the mass eigenstates
according to
\bea
\tilde{d}_{\!R_{\,{\scriptstyle i}}}&=&\!\!(V_d^{\!R0\,\dagger})_{\mbox{}_
{\!{\scriptstyle ij}}}\,
(\Gamma_{\!dR}^{\dagger})_{\mbox{}_{{\scriptstyle j\alpha}}}\,
\tilde{d}_{\alpha} \label{dR} \\
\tilde{d}_{\!L_{\,{\scriptstyle i}}}&=&\!\!(V_d^{\!L0\,\dagger})_{\mbox{}_
{\!{\scriptstyle ij}}}\,
(\Gamma_{\!dL}^{\dagger})_{\mbox{}_{{\scriptstyle j\alpha}}}\,
\tilde{d}_{\alpha}\;. \label{dL}
\eea
As indicated in these equations, the ${V}$ matrices rotate squarks the same
way as they do with quarks. The additional rotations are then performed by
the 6x3 matrices $\Gamma_{dL,R}$ .
Rules for the Feynman diagrams using this notation can be found in \cite{BBMR}.
Note that the indices $i,\,j$,... denote generation indices 1,2,3 , the
greek letters denote squark indices 1 to 6 and that the implicit summation
over the repeating indices is assumed.
Diagram with the gluino and $d$-type squarks in the loop contributes to
the $quark$ self-energy matrix (amputated two-point function) as :
\bea
-i\Sigma=(-i\sqrt{2}g_3)^2\,C_2({\bf 3})\,\int\!\!\frac{d^dk}{(2\pi)^d}
\,(\!-\!)\,P_R<\tilde{g}\bar{\tilde{g}}>P_R\hspace{2 cm}\nonumber\\
V_d^{\!L0\,\dagger}\,\Gamma_{dL}^{\dagger}
<\tilde{d}\,\tilde{d}^{\dagger}>\,\Gamma_{dR}\,V^{\!R0}_d)\,
+\,P_L...P_L\:+\not\!p\,-terms\,.
\eea
Only the term with the two right-handed projectors
corrects the mass matrix. The term indicated as $P_L...P_L$
contains similar corrections to ${\bf m}^{\dagger}$.
To get to the formula (\ref{md}) in the text one performs the rotation
to Euclidean space and integrates out angular variables. The integral
measure $dk$ in (\ref{gammad}) stands for $k^2\,d(k^2)$ and the integration
limits are assumed to be zero and infinity. Note that in the main text
${\bf m}_d^0$ was appended to these equations in not a very
ellegant way , but that is for later convenience.

\vspace{4 mm}
\par
 $ii) \:\;chargino \;diagram$

\vspace{3 mm}
Quark-squark-chargino interaction that is relevant for this paper reads
\bea
{\cal L}_{int}\!\!\!\!&=&\!\!\!\!\!(\bar{d}\,P_R\,(V^{\!*\dagger})_{2A}
\tilde{\chi}_A^c)\,{\bf \lambda_u}\,\tilde{u}_R
+\tilde{u}^{\dagger}_L\,{\bf \lambda_d}((U^{\!*\dagger})_{2A}
\bar{\tilde{\chi}}_A^cP_R\,d\,) \nonumber\\
&\mbox{}&\!\!\!\!\!-g_2\,(\bar{d}\,P_R\,(V^{\!*\dagger})_{1A}
\tilde{\chi}_A^c)\,\tilde{u}_L\:+\:h.c.\;.
\eea
u squarks are rotated to their mass eigenstates in exactly the same way as
the d squarks above, defining the $\Gamma_{\!uR,L}$ matrices.
Contribution to the d quark self-energy from this interaction reads
\bea
&\mbox{}&\!\!\!\!\!\!\!\!\!\!\!
-i\Sigma=(i)^2\int\!\!\frac{d^dk}{(2\pi)^d}\,U_{A2}
V_{B2}\,P_R\!<\!\!\tilde{\chi}_A^c\bar{\tilde{\chi}}_B^c\!\!>\!P_R\,
{\bf \lambda_u}\,V^{\!R0\,\dagger}_u\,\Gamma_{\!uR}^{\dagger}\!
<\!\!\tilde{u}\,\tilde{u}^{\dagger}\!\!>\!\Gamma_{\!uL}\,V^{\!L0}_u\,{\bf
\lambda_d}-
\nonumber\\
&\mbox{}&\!\!\!\!\!\!\!\!\!
-U_{A2}V_{B1}P_R\!<\!\!\tilde{\chi}_A^c\bar{\tilde{\chi}}_B^c\!\!>\!P_R\,
g_2\,V^{\!L0\,\dagger}_u\,\Gamma_{\!uL}^{\dagger}
\!<\!\!\tilde{u}\,\tilde{u}^{\dagger}\!\!>\!\Gamma_{\!uL}\,V^{\!L0}_u
\,{\bf \lambda_d}+P_L...P_L+\not\!p\,terms\,.\nonumber
\eea
$U$ and $V$ diagonalize the chargino mass matrix. The fact that one of their
indices is 1 (2), traces back the wino (higgsino) interaction in the
quark-squark-chargino vertex of the loop.
Summation over $A,\!B$=1,2 is assumed. Explicit forms of the $U$ and $V$
matrices and further details about the notation can be found in ref.\cite{HK}.
In order to derive the equation (\ref{md}) one has to use the relations
between the diagonalized and non-diagonalized mass and $\lambda$ matrices,
 briefly mentioned at the beginning of this appendix. The
vev of the scalar Higgs $H_d$ is added in order to pull out the mass
matrix on the r.h.s. for future convenience and when combined with
$tan\beta$ (which is pulled out into the $\epsilon$) it yields $(v_u)^{-1}$
in the final expression (\ref{gammau}) given in the text.

\vspace{.2in}

\noindent
{\bf\large Acknowledgements}

STP thanks Marek Olechowski for useful discussions.

\end{document}